\documentstyle[pre,aps,epsf,preprint]{revtex}
\draft
\begin{document}

\newcommand{\bm}{\bibitem}
\newcommand{\bgea}{\begin{eqnarray}}
\newcommand{\ndea}{\end{eqnarray}}
\newcommand{\bge}{\begin{equation}}
\newcommand{\nde}{\end{equation}}
\newcommand{\lbl}{\label}
\newcommand{\rf}[1]{(\ref{#1})}
\newcommand{\cc}{\mathop{\rm c.c.}\nolimits}
\newcommand{\hf}{\frac{1}{2}}
\newcommand{\rb}{Rayleigh-B\'{e}nard }
\newcommand{\bb}[1]{{\bf #1}}
\newcommand{\rr}{\bb r}
\newcommand{\vfi}{\varphi}
\newcommand{\prtt}{\partial_t}
\newcommand{\prtr}{\partial_r}
\newcommand{\prtf}{\partial_\phi}
\newcommand{\prtpl}{\partial_{\|}}
\newcommand{\prtpp}{\partial_\perp}
\newcommand{\prtR}{\partial_R}
\newcommand{\prtT}{\partial_T}
\newcommand{\prtF}{\partial_\Phi}
\newcommand{\ob}{Oberbeck-Boussinesq }
\newcommand{\omrm}{\omega_m R_m}
\newcommand{\vna}{\bb \nabla}
\newcommand{\lpl}{{\bb \nabla}^2}
\newlength{\absindent}
\setlength{\absindent}{\parindent}
\newcommand{\absind}{\hspace{\absindent}}
%
%
%
%
%
%
%
\title{Dynamical Properties of Multi-Armed Global Spirals in
	Rayleigh-B\'{e}nard Convection}
\author{Xiao-jun Li$^1$, Haowen Xi$^2$ and J. D. Gunton$^1$}
\address{$^1$Department of Physics,
	Lehigh University, Bethlehem, Pennsylvania 18015}
\address{$^2$Department of Physics and Astronomy,
	Bowling Green State University,
	Bowling Green, Ohio 43403}
\date{Submitted 20 May 1996}
\maketitle
\begin{abstract}
	Explicit formulas for the rotation frequency and the
long-wavenumber diffusion
coefficients of global spirals with $m$ arms
in \rb convection are obtained. Global spirals
and parallel rolls share exactly the same Eckhaus, zigzag and skewed-varicose
instability boundaries.
Global spirals seem not to have a characteristic frequency $\omega_m$ or
a typical size $R_m$, but
their product $\omrm$ is a constant under given experimental
conditions.
The ratio $R_i/R_j$ of the radii of any two dislocations ($R_i$, $R_j$)
inside
a multi-armed spiral is also predicted to be
constant. Some of these results have been
tested by our numerical work.
\end{abstract}




	Global spirals and {\it spiral defect chaos} (SDC)
as intrinsic patterns have been experimentally observed
recently in \rb convection (RBC) \cite{5,7,12}.
These observations
were rather surprising because they
were carried out in a parametric region where the familiar parallel-roll
pattern should be
stable \cite{4}.  An explanation for the unexpected presence of global
spirals or SDC,
in place of parallel rolls, is still to be provided.
So far  theoretical attempts on understanding these
intriguing patterns  rely heavily on numerical solutions of
either the
generalized Swift-Hohenberg (GSH) model \cite{8,9,10,14} or the truncated
Navier-Stokes equations governing the fluid dynamics \cite{11}. Although these
numerical studies have reproduced experimental results qualitatively and
quantitatively,
very limited theoretical insights have been obtained.
While the formation of SDC in the system
has since received considerable attention
\cite{10,11},
little effort has been given to determining
the essential properties of a {\it single}
spiral. It is far from clear
whether a global spiral has a characteristic rotation frequency
$\omega_m$ or a typical size $R_m$ \cite{2}.
The knowledge of those properties, we believe, is necessary in order
to describe the much more complicated SDC.

	Recently major progress
was made by Cross and Tu (CT) in this front \cite{14}.
Applying the {\it phase dynamics}  method developed earlier
in studying pattern formations in non-equilibrium systems \cite{17,20},
CT considered the
dynamics of a spiral as the balance of two competitive motions:
a radial phase-drifting of the rolls and an azimuthal
climbing of the dislocation  \cite{14}.
CT's results also imply  that $\omega_m R_m$
is a constant under given experimental conditions, but they
did not give an explicit expression for $\omega_m R_m$.
Furthermore,
CT demonstrated qualitatively that the rotation frequency
$\omega_m$ of a spiral is not directly related
to {\it mean flow}, which is induced by
distortions of the  convective rolls \cite{15},
although mean flow is necessary
for the formation of the rotating spiral \cite{8,9,10,11}.

	In this paper we focus on dynamical properties of {\it global} spirals.
We extend CT's results in two respects: We first
make a one-mode approximation for spirals \cite{23} which leads to  an
explicit expression for $\omega_m R_m$; we also separate the phase fluctuations
from the stable phase-drifting and calculate
the phase diffusion coefficients.
We test some of these
formulas by our numerical solutions.
Our results make it possible to discuss
the dynamical properties of global spirals in detail. We predict that,
inside a
stable multi-armed global spiral, the ratio of
the radii of two dislocations
is a constant under given experimental conditions.
We also find that global spirals, concentric rings and parallel rolls
have exactly the same Eckhaus, zigzag and skewed-varicose instability
boundaries. Presumably there is a competition among the various attractors
corresponding to these states, the nature of which requires further
theoretical study.

	To be concrete, we base our calculations on the two-dimensional
generalized Swift-Hohenberg (GSH) model for RBC
\cite{15}, which has been proven very successful in characterizing
convective patterns under quite broad conditions \cite{1}.
Numerical solutions of the GSH model not only reproduce
both global spirals and SDC but also resemble experimental results reasonably
well
\cite{8,9,10}.  In this model, the convective patterns are determined
completely by an {\it order parameter} $\psi(\rr,t)$
in two-dimensional space $\rr$, which satisfies \cite{15}
\bge
	\prtt \psi + \bb U \cdot \bb \nabla \psi
	= \left[\epsilon -({\bb \nabla}^2 + 1)^2 \right] \psi
		-g \psi^3 + g_3 (\bb \nabla \psi)^2 {\bb \nabla}^2 \psi,
				\lbl{gsh}
\nde
where $\bb U$ is the mean flow velocity given by
$\bb U = \vna \zeta \times {\bb e}_z$ while \cite{15}
\bge
	\left[\prtt -\sigma (\lpl - c^2)\right] \lpl \zeta
		= g_m {\bb e}_z \cdot \left[ \vna(\lpl \psi)
		\times \vna \psi\right]. \lbl{mf}
\nde
In the GSH equations, the reduced Rayleigh number
$\epsilon = 2.7824 \epsilon_{\rm expt}$, where
$\epsilon_{\rm expt} \equiv (Ra/{Ra}_c) -1$
is the control parameter \cite{8,10},
in which  $Ra$ and ${Ra}_c$ are
the Rayleigh number and its critical value at onset \cite{1}. Other
parameters $g$, $g_3$, $c^2$, $g_m$ and the fluid Prandtl number
$\sigma$ model the properties of the system
and are all non-negative.
For simplicity,
we only consider \ob fluids here \cite{26}.

	A one-mode approximation has been used in studying spirals in chemical
reaction-diffusion systems \cite{23}. We now apply the same approximation
for stable global spirals in RBC. Using polar coordinates $(r, \vfi)$, one can
approximate a global-spiral solution by \cite{26}
\bge
	\psi(\rr ,t) = \hf \left[A_m(r) e^{i \theta_m} + \cc\right]
		+ O(A_m^3), \lbl{mode}
\nde
where $\theta_m=k_m(r) r +m \vfi - |m| \omega_m t$ with
$m$  the number of spiral arms near the core. In general,
the amplitude $A_m(r)$   and  the wave number
$k_m(r)$  should depend on $r$.  The rotation frequency $\omega_m$, however,
must be independent of $r$ for a stable spiral.
Here we adopt the following
conventions: $k_m(r) >0$, $m >0(<0)$ if the spiral is
right(left)-handed and $\omega_m >0(<0)$ if the spiral rotates in the
same (opposite) direction of chirality.
Experiments \cite{5,12} showed that
a multi-armed spiral usually has dislocations at different radii. Across
each of these radii,
the number of arms decreases or increases (depending on whether
$m>0$ or $m<0$)
by the number of dislocations on the corresponding radius.
For mathematical
simplicity, we denote each  of these radii, for example  $R_i$,
by the number of non-terminating
arms, say $i$, in its inside vicinity: See Fig. \ref{fig}
for a $m=3$ global spiral.
Then
a multi-armed spiral with dislocations at different radii
can also be decribed by the above solution, provided that
$m$ is replaced by $i$ for each corresponding region $R_j < r \le R_i$.
The amplitude and the
wave number should be continuous across each boundary $r=R_i$.
The frequency, on the hand, must be
a constant for all regions.  For $r>R_1$, only
concentric rings (a target state) exist which corresponds to $i=0$.
With this understanding,
our results below can be directly extended, by replacing $m$ with every
possible $i$, to multi-armed spirals found in experiments.

	Phase equation \cite{17,20}
describes slow variations of convective rolls from their perfect
pattern.
Assuming $R_m \gg 1$, i.e.,
$\eta_m^2 \equiv 1/R_m \ll 1$ for a global spiral, one can then
introduce  {\it slow} scales
$R \equiv \eta_m^2 r$,  $\Phi \equiv \vfi$ and $T \equiv\eta_m^4 t$,
and a {\it slow} phase variable
$\Theta_m(\bb R, T) \equiv \eta_m^2 \theta_m(\rr,t) + |m| \omrm T$.
Here the phase fluctuations $\Theta_m(\bb R,T)$ have been explicitly separated
from the stable phase-drifting $|m| \omrm T$. A similar separation has been
used in studying chemical waves with steady velocity \cite{1}. Now the
local wave number can be defined as \cite{17}
\bge
	{\bb q}_m \equiv \vna_{\rr} \theta_m(\rr,t)
		= \vna_{\bb R} \Theta_m(\bb R,T) = q_m {\bb e}_r +O(\eta_m^2).
			\lbl{wn}
\nde
Inserting the
spiral solution \rf{mode} into the GSH equation \rf{gsh}, one may then
match the result to order  $\eta_m$. Since
the calculations are carried out in exactly the same way
as in Ref. \cite{17}, we skip details here but only write down
the final results. From the zeroth order of $\eta_m$,
one finds that the amplitude
is slaved by the wave number and is given by \cite{25}
\bge
	|A_m|^2 = 4 \frac{\epsilon -(1-q_m^2)^2}{3 g + g_3 q_m^4}.
		\lbl{am}
\nde
{}From the second order of $\eta_m$, one obtains essentially the phase equation
\bgea
	\prtT \Theta_m &=& |m| \omrm
		-|A_m|^{-2} \vna_{\bb R} \cdot
		\left[B(q_m) |A_m|^2 {\bb q}_m \right] \nonumber \\
		&&-{\bb U}' \cdot {\bb q}_m   - g_3 q^{1/2}_m {\bb q}_m \cdot
		\vna_{\bb R} \left[q_m^{3/2}|A_m|^2\right],
		\lbl{pheq}
\ndea
where
\bge
	B(q_m) = 2 (1-q_m^2) -\frac{3}{4} g_3 q_m^2 |A_m|^2, \lbl{bfn}
\nde
while ${\bb U}' = \vna_{\bb R} \langle\zeta\rangle_{\theta_m}
\times {\bb e}_z$ with $\langle \cdots \rangle_{\theta_m}$ for the phase
average in $\theta_m$ and,
with $g'_m=g_m/c^2 \sigma$,
\bge
	\lpl_{\bb R} \langle\zeta\rangle_{\theta_m}
		= - \hf g'_m {\bb e}_z \cdot {\bb q}_m \times \vna_{\bb R}
		\left[\vna_{\bb R} \cdot ({\bb q}_m |A_m|^2)\right].
			\lbl{mfph}
\nde
Although one may, in principle, convert Eq. \rf{pheq} into
the standard form as in Ref. \cite{17}, there is no need to do so here.

	To express the phase equation in the more familiar
diffusion equation form,
one needs to project the gradient operator $\vna$ into local coordinates.
In the light of Eq.~\rf{wn}, this can be easily achieved by $\prtpl=\prtR$
and $\prtpp=R^{-1}\prtF \simeq \prtF$ (since $R \simeq 1$ for $r \simeq R_m$
where the phase equation is valid).
Again recalling Eq.~\rf{wn}, one finds immediately that
$\vna_{\bb R} \cdot {\bb q}_m = \lpl_{\bb R} \Theta_m$. In  Cartesian
coordinates, this simply gives $\prtpl^2 \Theta_m + \prtpp^2 \Theta_m$ which
contributes to the diffusion of phase fluctuations. But
in a polar coordinate system,
an additional $R^{-1} \prtpl \Theta_m \simeq q_m$ term is also present
which, however, contributes to the stable phase-drifting $|m| \omrm T$!
{}From Eq.~\rf{pheq},
one gets via this $R^{-1}\prtpl \Theta_m \simeq q_m$ term that \cite{25}
\bge
	|m| \omrm = q_m B(q_m), \lbl{frq}
\nde
whose corrections are of order $\eta_m^2$.
Apparently this frequency of rotation
is generated by the curvature of convective
rolls although it appears independent of mean flow. For $m=0$, it naturally
leads to the ``wave number selection'' $q_0 = q_f$ with $B(q_f) = 0$
\cite{17,20}.

	Further algebra reduces Eq.~\rf{pheq} to
a phase diffusion equation \cite{17,22}
\bge
	\prtT \Theta_m = D_{\|}(q_m) \prtpl^2 \Theta_m
		+ D_{\perp}(q_m) \prtpp^2 \Theta_m 
		+D_\times(q_m) \vna^{-2}_{\bb R} \prtpl^2 \prtpp^2 \Theta_m,
			\lbl{diff}
\nde
in which the diffusion coefficients are \cite{25}
\bgea
	D_{\|}(q_m) &=& -\hf q_m \left[4 \frac{1-q_m^2}{|A_m|^2}
		-g_3 q^2_m\right] \frac{d}{d q_m}|A_m|^2
		-B(q_m) +4 q_m^2, \lbl{eck} \\
	D_\perp(q_m) &=& -B(q_m)+\hf g'_m q_m^2 |A_m|^2, \lbl{zz} \\
	D_\times(q_m) &=& \hf g'_m q^3_m \frac{d}{d q_m} |A_m|^2. \lbl{cross}
\ndea
This diffusion equation describes three
types of long-wave-length fluctuations \cite{17,22}:
Eckhaus ($D_{\|}$), zigzag  ($D_\perp$) and skewed-varicose ($D_{\rm sv}$)
with, for $D_\times < 0$,
\bge
	D_{{\rm sv}}(q_m)=\left[(D_{\|}-D_\perp)^2+2 (D_{\|}+D_\perp) D_\times
			+D_\times^2\right]/4 D_\times. \lbl{sv}
\nde
When all these diffusion coefficients are positive, global spirals are stable;
but when any one of them becomes negative, spirals lose their
stability against the corresponding fluctuations.

	One striking feature of this diffusion equation
is that all the functions
$D_{\|}$, $D_\perp$ and $D_{{\rm sv}}$ are {\it independent} of
$m$. Indeed they all agree with those of parallel rolls.
This means that global spirals, concentric rings and parallel rolls
share exactly
the same Eckhaus, zigzag and skewed-varicose instability boundaries.
So the fact that the measured wave numbers
of stable spirals are inside the stable region of
parallel rolls \cite{5} is not
surprising but necessary.
Considering that the system is
{\it non-potential} and that the only difference between spirals of different
number of arms is $q_m$,
a theoretical understanding of spiral-to-target or spiral-to-spiral
transitions could be subtle. There are, however, some shortcomings in
our analysis. The core instability \cite{14} is omitted here.
Short-wavelength fluctuations \cite{4,22} have also been neglected, but
these seem irrelevant in the transition  between global spirals of different
value of $m$ \cite{5,12}. In any event, it is clear that further theoretical
work on this issue is necessary.

	We now concentrate on the rotation frequency of the global spirals.
One sees easily from Eq.~\rf{frq} that
$\omrm$ is a constant under given experimental conditions.
This has been implied by CT \cite{14} and verified by experiments \cite{13} and
our numerical solutions: see Table~\ref{table}.
The quantitative value of $\omrm$ apparently depends on $q_m$. However,
except for $m=0$, one cannot determine $q_m$ merely from the
phase equation. This
problem has been emphasized by CT \cite{14}. They found that it is
essential to take the intrinsic defect of the spiral, i.e., the dislocation
defect, into account.  A spiral can be
stable only if the phase-drifting of the rolls is balanced by the
climbing of the dislocation, i.e., $\omrm = v_d(q_m)$, where $v_d(q_m)$ is the
climbing velocity of the dislocation. This, together with Eq.~\rf{frq},
selects the wave number for a stable spiral. Unfortunately, an explicit
analytic formula of $v_d(q_m)$ seems intractable \cite{21}. Thus an
accurate evaluation of $q_m$ seems beyond reach.

	One may still
make the following approximations for small $\epsilon$ \cite{14},
\bge
	|m| \omrm \simeq \alpha (q_f -q_m), \quad
	v_d(q_m) \simeq \beta (q_m-q_d), \lbl{clmb}
\nde
where $\alpha,\, \beta > 0$ and $q_d$ is defined by $v_d(q_d) = 0$.
For $g_3 = 0$, we get from Eq.~\rf{frq} $q_f = 1$ and $\alpha =4$, while
for $g_3 >0$, we get $q_f \approx 1- \gamma_f \epsilon$ and
$\alpha \approx  4(1+ \tilde{\alpha} \epsilon)$
with $\gamma_f = 3 g_3/4(3 g+g_3)$ and
$\tilde{\alpha}=6 \gamma_f+\frac{32}{3} \gamma_f^2$. Similarly, one may write
$q_d \approx 1 - \gamma_d \epsilon$ and
$\beta \approx \beta_0 (1+\tilde{\beta} \epsilon)$.
Under these
approximations, the selected wave number of a stable
spiral is fully determined by $\omrm = v_d$, which gives, with
$\gamma_m = (4 \gamma_f+|m| \beta_0 \gamma_d)/(4+|m| \beta_0)$,
\bge
	q_m \simeq (\alpha q_f +|m| \beta q_d)/(\alpha+|m| \beta)
		\approx 1 - \gamma_m \epsilon. \lbl{slctwn}
\nde
Then, from Eq.~\rf{frq}, one finds for $m \neq 0$ that
\bge
	\omrm \simeq \frac{\alpha \beta}{\alpha+|m| \beta} (q_f-q_d)
		\approx \frac{4 \beta_0 (\gamma_f-\gamma_d) \epsilon}
		{4+|m| \beta_0} \,.
			\lbl{apprxfrq}
\nde
For a multi-armed spiral with dislocations at different radii, since
the frequency of rotation is a constant over the whole spiral, the ratio
$R_i/R_j = (\alpha+|j| \beta)/(\alpha+|i| \beta)$
is fixed under given experimental conditions. (The definition
of $R_i$ and $R_j$ is given below Eq. \rf{mode}.)
This ratio
depends on only two experimentally measurable quantities $\alpha$ and
$\beta$ \cite{13}, which provides a strong test of our theory.

	To test the validity
of Eq. \rf{frq}, we have solved Eqs.~\rf{gsh}
and \rf{mf} numerically by the same method described in Ref.~\cite{8}, except
that we set up a global spiral as the initial condition. We use $g=1.0$,
$g_3=0$, $\sigma=1.0$,
$c^2=2.0$ and $g_m=10$ for our numerical solutions.
We use mesh points $N \times  N  = 512 \times  512$ and grid size
$\Delta  x = \Delta  y = \pi /8$ for aspect ratio (radius/thickness)
$\Gamma=32$. For simplicity, we set all radii of dislocations inside a
multi-armed spiral to be equal.
For $\epsilon =0.3$, a one-armed spiral shrinks rapidly
into concentric rings.
But for $\epsilon =0.5$,
one-armed, two-armed and three-armed spirals all
stabilize with three given sizes $R_m/L = 0.45, 0.55, 0.65$ where $L=32 \pi$ is
the radius of the cell. We then measure
the rotation frequency and, via Eq. \rf{frq}, calculate the corresponding
wave number $q_m$. The results are listed in Table~\ref{table}.
Although $R_m$
varies by about $50\%$, the product $\omrm$ is
found to be a constant within $5\%$.
Also from Eq. \rf{apprxfrq} and the measured values of $\omrm$, we find that
$\beta = 0.73 \pm 0.08$ and $q_d = 0.880 \pm 0.001$. In comparison, one gets
$q_d \simeq 0.905$ from a formula in Ref.~\cite{21}.
Furthermore, although global spirals are set up with three different sizes,
they are all found to be stable
within the time of our computer simulation, which runs
about $2000$ vertical diffusion times for each case.
So it seems that global spirals do not have a typical size.
Consequently,
the frequency of rotation may also be non-unique.

	Finally we make a comment on the role played by mean flow in the
dynamics of spirals. Evidently Eq.~\rf{frq} does not explicitly
depend on $g_m$. This has been first observed by CT \cite{14}, who
hence assign a ``secondary'' role to mean flow. Nevertheless mean flow plays
a subtle role in determining
the selected wave number  \rf{slctwn} for {\it stable} spirals.
Indeed the value of $q_f -q_d$ in Eq.~\rf{apprxfrq} is very sensitive to $g_m$
\cite{21}. So $\omrm$ also has a sensitive $g_m$ dependence.
Furthermore, one must have $D_\perp \ge 0$ for stable spirals. This, from
Eqs.~\rf{zz} and \rf{frq}, gives a constraint on the allowed
frequency. For $g_m=0$, only
$\omrm \le 0$ is permitted. Now if $\omrm \ge 0$ is also necessary to avoid
``unwinding'' \cite{14}, stable spirals must be stationary. Recalling
Eq.~\rf{apprxfrq}, a stationary spiral is possible only if $q_f = q_d$.
For $g_3 =0$ and $g_m =0$,
the relation $q_f = q_d$ indeed holds \cite{21} and a stationary global
spiral has been found \cite{8}. But for more realistic $g_3 >0$,
these two wave numbers are in general unequal. So
a finite $g'_m$ is needed to observe any stable spiral, which
might suggest why a low Prandtl number ($g'_m \sim 1/\sigma$)
is necessary in experiments \cite{5,7,12}.

	In summary, we have calculated the rotation frequency and the
long-wavelength
diffusion coefficients of global spirals. We find that global spirals have
exactly the same Eckhaus, zigzag and skewed-varicose instability boundaries
as parallel rolls and concentric rings.
So a transition between these patterns presumably involves a competition
among their various attractors.
Although global spirals seem not to have
a characteristic frequency or a typical size, the product of them
is a constant under given experimental conditions. The ratio of the
radii of any two dislocations inside a multi-armed spiral is also predicted
to be constant. Some of these results have been tested by our numerical
solutions.
Nevertheless, to fully understand the intriguing global-spiral pattern,
an analysis of the core instability and a theory
describing the spiral-to-target
transition will be necessary.

	XJL and JDG are grateful to the National Science Foundation for
support (under Grant No. DMR-9596202). Numerical calculations reported
here are carried out on the Cray-C90 at the Pittsburgh
Supercomputing  Center and Cray-YMP8 at Ohio Supercomputer Center.

%
%
%
%
%
%
%

\begin{figure}
        \makebox
		{\epsfbox{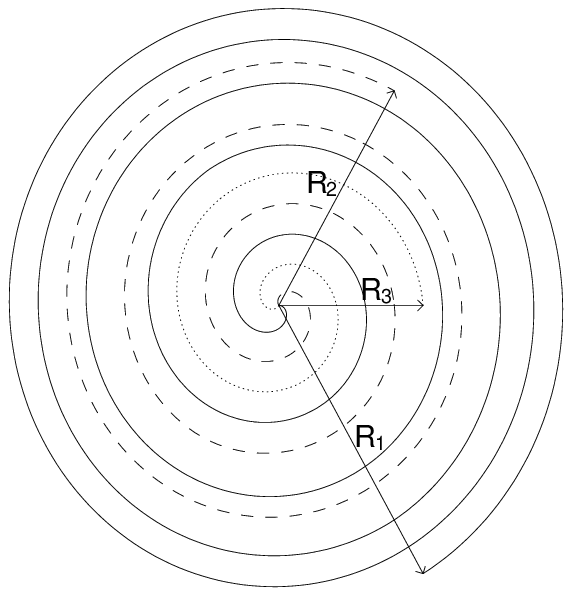}}
        \caption{Three-armed global spiral with dislocations at
			$R_3$, $R_2$ and $R_1$ ($R_3 < R_2 < R_1$).
			The three arms are
			represented by the solid, dashed and dotted lines.}
                \lbl{fig}
\end{figure}

\begin{table}
\caption{Measured values of $\omrm$ of global spirals with $m=1,\, 2,\, 3$
	from numerical solutions. Global spirals are set up with three
	different sizes $R_m/L = 0.45,\, 0.55,\, 0.65$ where $L=32 \pi$ is
	the radius of the cell.
	The wave number $q_m$ is calculated via
	Eq. (9), which is within the uncertainty of
	direct measurements $q_m =1.00 \pm 0.05$.}
\begin{tabular}{cccc}
$m$ & $R_m/32 \pi$ & $\omrm$ & $q_m$ \\ \hline
$1$ & $0.45$ & $0.0750$ & $0.981$ \\
	& $0.55$ & $0.0742$ & $0.981$ \\
	& $0.65$ & $0.0734$ & $0.981$ \\
$2$ & $0.45$ & $0.0619$ & $0.967$ \\
	& $0.55$ & $0.0630$ & $0.967$ \\
	& $0.65$ & $0.0658$ & $0.965$ \\
$3$ & $0.45$ & $0.0574$ & $0.954$ \\
	& $0.55$ & $0.0568$ & $0.954$ \\
	& $0.65$ & $0.0556$ & $0.955$ \\
\end{tabular}
\lbl{table}
\end{table}




\begin{references}
\bm{1} M.~C.~Cross and
	P.~C.~Hohenberg, Rev. Mod. Phys. {\bf 65}, 851 (1993).
\bm{2} G.~Ahlers, {\it Over Two Decades of Pattern Formation,
	A Personal Perspective} [preprint].
\bm{4} F.~H.~Busse, Rep. Prog. Phys. {\bf 41}, 1929 (1978).
\bm{5} E. Bodenschatz {\it et al.}, Phys. Rev. Lett. {\bf 67}, 3078 (1991);
	E. Bodenschatz {\it et al.}, Physica D {\bf 61}, 77 (1992).
\bm{7} S.~W.~Morris {\it et al.}, Phys. Rev. Lett. {\bf 71}, 2026 (1993).
\bm{8} H.-W.~Xi, J.~Vi\~{n}als and J.~D.~Gunton, Phys. Rev. A {\bf 46},
	R4483 (1992); H.-W.~Xi, J.~D.~Gunton and J.~Vi\~{n}als,
	Phys. Rev. E {\bf 47}, R2987 (1993).
\bm{9} M.~Bestehorn {\it et al.}, Phys. Lett. A {\bf 174}, 48 (1993).
\bm{10} H.-W.~Xi, J.~D.~Gunton and J.~Vi\~{n}als, Phys. Rev. Lett. {\bf 71},
	2030 (1993); H.-W.~Xi and J.~D.~Gunton, Phys. Rev. E {\bf 52}, 4963
	(1995).
\bm{11} W.~Decker, W.~Pesch and A.~Weber, Phys. Rev. Lett. {\bf 73}, 648
	(1994).
\bm{12} B. B. Plapp and E.~Bodenschatz [preprint].
\bm{13}
	E.~Bodenschatz and B. B. Plapp, Bull. Am. Phys. Soc. {\bf 41}(1),
	701 [Q18.8] (1996).
\bm{14} M.~C.~Cross and Y.~Tu, Phys. Rev. Lett. {\bf 75}, 834 (1995).
\bm{15} J.~Swift and P.~C.~Hohenberg, Phys. Rev. A {\bf 15}, 319 (1977);
	E.~D. Siggia and A.~Zippelius, Phys. Rev. Lett. {\bf 47}, 835 (1981);
	P.~Mannevill, J. Phys. (Paris) {\bf 44}, 759 (1983).
\bm{17} M.~C.~Cross and A.~C.~Newell, Physica D {\bf 10}, 299 (19984);
\bm{20} Y.~Pomeau and P.~Manneville, J. Phys. (Paris) {\bf 40}, L609 (1979);
	{\it ibid} {\bf 42}, 1067 (1981).
\bm{21} Y.~Pomeau, S.~Zaleski and P.~Manneville, Phys. Rev. A {\bf 27}, 2710
	(1983); G. Tesauro and M. C. Cross, {\it ibid} {\bf 34}, 1363 (1986).
\bm{22} H.~S.~Greenside and M.~C.~Cross, Phys. Rev. A {\bf 31}, 2492 (1985).
\bm{23} S.~Koga, Prog. Theor. Phys. {\bf 67}, 164 (1982).
\bm{25} Following Pomeau and Manneville \cite{20}, an alternative method is
	to impose a small fluctuation $\phi(x,y,t)$ to the phase $\theta_m$
	in Eqs.~\rf{gsh} and \rf{mode}, and match the gradients
	$\nabla^n \phi$ order by order. This leads to the same results
	as cited in the text.
\bm{26} It is straight-forward to extend our calculation to the non-\ob
	case in which a $-g_2 \psi^2$ term will be added to Eq.~\rf{gsh}.
	Consequently, in addition to mode \rf{mode}, two extra
	modes $\hf (A_{m,0}+A_{m,2} e^{2 i \theta_m} + \cc)$
	are needed. Since
	these two extra modes are generated by the $\psi^2$ term, their
	amplitudes are both of order $A_m^2$.
\end{references}
\end{document}